\documentclass[conference,10pt]{IEEEtran}
\usepackage{pslatex}
\usepackage{graphicx}
\usepackage{pst-node}
\usepackage{textcmds}
\usepackage{textcmds}

\onecolumn

\begin{document}

\title{\Large SANA - Network Protection through artificial Immunity}

\author{
\authorblockN{Michael Hilker \& Christoph Schommer}
\authorblockA{{\small University of Luxembourg, Campus Kirchberg}\\
{\small Dept. of Computer Science and Communication}\\
{\small 6, Rue Richard Coudenhove-Kalergi, L-1359 Luxembourg}\\
{\small Email: \{michael.hilker, christoph.schommer\}@uni.lu}}}

\maketitle

\thispagestyle{plain}

\begin{abstract}
Current network protection systems use a collection of intelligent components - e.g. classifiers or rule-based firewall systems to detect intrusions and anomalies and to secure a network against viruses, worms, or trojans. However, these network systems rely on individuality and support an architecture with less collaborative work of the protection components. They give less administration support for maintenance, but offer a large number of individual single points of failures - an ideal situation for network attacks to succeed. In this work, we discuss the required features, the performance, and the problems of a distributed protection system called {\it SANA}. It consists of a cooperative architecture, it is motivated by the human immune system, where the components correspond to artificial immune cells that are connected for their collaborative work. SANA promises a better protection against intruders than common known protection systems through an adaptive self-management while keeping the resources efficiently by an intelligent reduction of redundant tasks. We introduce a library of several novel and common used protection components and evaluate the performance of SANA  by a proof-of-concept implementation.
\end{abstract}


\IEEEpeerreviewmaketitle

% =================================================================
\section{Introduction}
\label{secIntroduction}
% =================================================================
A network protection system is a system that tries to protect the network and its nodes against intrusions and attacks. It consists of several components of different granularity, ranging from parameter changes in sub-components, nodes, or packet filters to powerful antivirus software, firewalls, and intrusion detection systems in important network nodes \cite{Roe99}. Each protecting component contains the workflow, tasks performed by this component, and the knowledge to perform the tasks so that the protection system converges to its goals. The protection system defines the cooperation between the different protection components and the maintenance- and update workflows. The performance of the protection system measures how secure the network is against intrusions; the performance originates out of the performance of all protection components.

There exist different characteristics in order to evaluate and compare different protection systems \cite{Deb99}:
\begin{itemize}
\item The system has to be complete; this means that the system protects all nodes against intrusions. 
\item The provided resources has to be used efficiently.
\item The protection system has to secure the network and all nodes as well as it should not influence the normal production of the network.
\item The installation, maintenance, and update workflows has to be easy and fast to perform. The system has to ensure that the different protection system work properly and are up-to-date.
\item The system has to adapt to the current situation in the network and it has to be adaptive in order to identify modified, mutated and/or novel attacks.
\end{itemize}

An important characteristics of protection systems is that they should be extendable and cope with upcoming - more and more complex as well as intelligent - intrusions. For example, the {\verb ZMIST } virus remains hidden in host programs; common-used detection techniques performing pattern matching have serious problems to identify it. The {\verb EVOL } virus hides itself through swapping instructions in existing programs. Thus, we demand for a protection system to use an adaptive information management having protection components collaborated; we demand for protection systems to be easily extendable towards and permanent protection update.

The section \ref{secSANA} introduces a novel framework for protection system, which is motivated by the human immune system \cite{Jan04} and implements an artificial immune system \cite{Aic04,Dec02,Hof99,Hof00} where the artificial immune system is classified as a complex adaptive system without a centralised system \cite{Var02}. It provides the features and characteristics explained above and provides a dynamic, efficient, and adaptive security environment in which common used and novel approaches for network security are combined \cite{Hil06c}.

% =================================================================
\section{What does exist?}
\label{secCurrentSituation}
% =================================================================
Current protection is done by architectures having different components. In each node, antivirus software and firewall are installed; they observe file access, active processes, and the network packet headers for possible attacks. Furthermore, on hubs, switches, and routers run header checking packet filters in order to define a network policy describing which routings, protocols, and ports are allowed to use. Important nodes like e.g. the internet gateway and the email server are secured using intrusion detection systems, which check each packet completely and observe the whole node in order to identify intrusions. 

These systems must be installed and configured in each node manually. The update of each protection component works as follows: each component asks regularly a central update server whether there is an update; if yes, then it downloads it. The antivirus software and the firewall ask then the user when a suspicious event occurs; the user takes over the decision and proceeds. This is often a risk, since the user is mostly not an expert. Additionally, the antivirus software, firewalls, packet filters, and intrusion detection systems use a log system in order to inform the administrator about different event. The administrator has to analyse these entire messages and react to it. Due to the enormous number of messages is it not possible that the administrator can analyse each message properly and the intrusions have the possibility to stay undetected. 

The different protection components are not connected. Thus, each component works on its own and collaborative work does not exist. This leads primarily to redundant checks because e.g. the firewall and intrusion detection system in a node check the packet for the same characteristics and consequently resources are wasted. However, more important is that the different information gathered through the different protection components are not combined in order to identify intrusions or even to identify abnormal or suspicious behaviour. Lastly, the protection system does not check itself whether each component is up-to-date and works properly, which leads to the situation that different protection components work with limited performance. An example for this problem is that the update system of an antivirus software does not work properly and e.g. the antivirus software is no longer updated and does not identify the newest intrusions; this is a problem because this node is a risk for the whole network.

Some approaches for a distributed protection system exist, e.g. using an artificial immune system for network security \cite{Aic04}. However, these systems mostly do not use a fully distributed approach, i.e. there are some centralised components which are critical for the overall performance of the system, the system is mostly not adaptive, and the system does not use small entities as artificial cells and more the multi-agent system approach \cite{Bel99} using heavy agents with few mobility, lots of internal information, and lots of knowledge about the current situation. Furthermore, most system do not use an approach where many artificial cells have to communicate and collaborate. Furthermore, these systems are far away from production; this framework intends to narrow the gap between the distributed protection systems of academic research and production.

In the next section, a novel framework for a distributed, integrated, and dynamic protection system is introduced in which common used and novel approaches for network security are combined and which tries to solve the limitations of existing protection systems. 

% =================================================================
\section{Artificial Immune System SANA}
\label{secSANA}
% =================================================================
The protection system SANA \cite{Hil06c} is a novel framework with a library of different protection components using a more sophisticated organisation. SANA tries to cope with upcoming attacks that increase continuously adaption and intelligence. SANA's administration and maintenance is simplified, it works mainly autonomously. The system does not use centralised protection components, which are single points of failure. 

% =================================================================
\subsection{Security Environment}
% =================================================================
The security environment is installed in each node and provides an environment that is used by each protection component. The environment ensures the access to resources as storage, memory, CPU, and network as well as ensures that each protection component is recognised when a certain event occurs. Therefore, each protection component is installed in the security environment and registered for certain events, e.g. arriving of a packet or access to a file. Thus, the protection components are independent from the underlying hardware platform as well as the network protocols and configuration - communication middleware between network protocols and protection components. Furthermore, the components are also independent from the operating system because the security environment handles the access to the resources.

The security environment also provides checks testing the protection components if they are allowed to access the resources and if they intend to perform properly. This ensures that not properly working components are removed from the network and cannot access resources anymore.

% =================================================================
\subsection{Artificial Cells and other Protection Components}
% =================================================================
The protection system consists of several protection components, which perform the tasks required for network security. These protection components are both common used protection components like antivirus software, firewall, packet filter, and IDS in order to benefit from their performance as well as novel approaches of protection components, e.g. the artificial cells. The common used protection components are installed in the security environment and perform their tasks.

The artificial cells are the novel protection components. These cells are dynamic, highly specialised, lightweighted, and mobile so that the system is dynamic and hard to attack. The number of different types of artificial cells is enormous because the tasks required for network security are split in small pieces so that a breakdown of one artificial cell does not break down the whole system. Furthermore, each type of artificial cell is several times present in the network so that redundancies occur. Examples for artificial cells are introduced in \cite{Gre05, Hil06b}.

Examples for artificial cells are e.g. cells that perform intrusion detection in checking each network packet for a certain pattern; e.g. type of artificial cell uses exactly one rule to identify one certain intrusion. Otherwise, there are types, which reuse the information gathered by other protection components in order to identify infected nodes, and types, which check the status of nodes and protection components in order to identify not proper working systems. However, it is possible to define artificial cells with lots of different possible tasks, workflows, and behaviours. 

The artificial cells shutdown over time and systems in the network generate continuously new cells in order to keep the population up-to-date. With this workflow, novel approaches and techniques for network security can be introduced. The communication and collaboration of artificial cells is discussed in the next section. 

% =================================================================
\subsection{Artificial Cell Communication}
% =================================================================
The artificial cell communication models the cell communication of the human body in order to provide a communication and collaboration protocol for the artificial cells and other protection components. Therefore, the term {\it artificial substance} is introduced, which are used for message exchanges and models the behaviour of different substances of the human body, e.g. the cytokines and hormones \cite{Jan04}. Each artificial substance contains the message and a header with the parameters {\verb hops-to-go } and {\verb time-to-live } describing the distribution area of the artificial substance. Each node manages the routing of the substances where each node has a set of nodes; each substance is sent to these nodes when the distribution area is not reached; this set is changed according to the current situation in the network. 

Two types of nodes are additional introduced. The artificial lymph nodes supply the artificial cells with additional information, collect status information, and respond to artificial substances. This response can be to inform the administrator or to release additional artificial cells in order to perform certain tasks. The artificial lymph nodes are installed in each network equipment like hubs, routers, and switches. The central nativity and training stations (CNTS) generate continuously artificial cells and release these cells to the network in order to keep the population up-to-date. Furthermore, the CNTS collect information about the current situation in the network in order to provide on demand a status snapshot for the administrator and to adapt the generation of the cells to the current situation. Novel techniques and knowledge can be introduced in the network through releasing novel artificial cells. 

The artificial cell communication works as follows that one component wants to send a message, packs this message in an artificial substance, defines the distribution area, and gives it to the network. In each node, the artificial substance is presented to all protection components and the right components receive the message. The identification of the right receivers is done using artificial receptors. These receptors are a public/private key pair describing the type and status of the artificial cells, protection components, and artificial substances; all artificial cells, protection components, and artificial substances contain several artificial receptors. Only when an artificial cell or protection component can authorise itself using the right keys of the artificial receptors, it will receive the message stored in the artificial substance. The protection components receiving the message respond to it and the node sends the substance to all next nodes and the process repeats until the distribution area of the substance is reached. The artificial receptors are additionally used in order to secure the access to resources using authentication processes.

The advantages of the artificial cell communication are a distributed, efficient, and fail-safe protocol without a single point of failure. Furthermore, the protocol works fine for point to multi-point communication as used in network security where a component sends a message to all nearby components of a certain type. 

% =================================================================
\subsection{Self Management}
% =================================================================
The self management of the artificial immune system organises the numerous mobile artificial cells. Each protection component knows how much security it provides - a security value. Each node calculates - based on the security values of each component in the node - a security level; when this level falls below a certain threshold, it starts a notification process. This notification process attracts other nearby artificial cells in order to move to this node so that the security value is increased and the artificial cells in this node are more affine to stay than to move. However, the artificial cells still work autonomously. 

The self management increases the performance enormously because the artificial cells are properly distributed over all nodes and enough artificial cells can still keep moving in order to provide a dynamic protection system. 

% =================================================================
\subsection{Implementation}
% =================================================================
A platform-independent proof-of-concept implementation is done so that SANA can be compared with other approaches. The simulation runs on a network simulator implementing a packet oriented network - e.g. TCP/IP - where an adversarial stresses the network and the protection system using many packets with and without intrusions. Furthermore, the implementation can be easily extended and also common used protection systems as antivirus software, intrusion detection systems, and firewalls are implemented in order to evaluate the performance of SANA. Different simulations with real network attacks are installed and the performance is discussed in the next section. 

% =================================================================
\subsection{Performance and Results}
% =================================================================
The performance of SANA is more than acceptable. It performs better than common used protection systems, which are a collection of protection components, because SANA also uses the same common used protection components. Additionally, it contains the artificial cells providing a dynamic and adaptive part. Furthermore, SANA provides using the security environment an easy to administrate protection system that has many autonomous workflows. The warnings and alerts of each protection component are analysed and processed automatically by other protection components in order to adapt the behaviour - a more sophisticated information management. Therefore, different workflows are used: first, the information from the protection components are combined in order to identify infected nodes and abnormal behaviour; second, the infected nodes are disinfected and abnormal behaviour is observed; third, the warnings and alerts are sent to the nearby artificial cells which adapt their internal thresholds accordingly - implementation of the danger model \cite{Aic04a,Mat02}. Furthermore, the protection system is highly dynamic: the artificial cells move continuously and provide a hard to predict and to attack protection system. Furthermore, the system can be quickly extended with novel approaches for network security. 

In the simulations, real network attacks are modelled where e.g. different worms use the network to propagate and the aim of the worm is to infect as many nodes as possible. Two simulations are performed: the first with a common used protection system consisting of antivirus software, firewall, packet filter, and intrusion detection system, and a second with SANA. In all of our simulations SANA is more secure than the common used protection system. Furthermore, SANA adapts to the current situation because identification of infected nodes, not proper working components, and suspicious behaviour is detected and the problems are fixed automatically. 

Especially, when an intrusion uses the network for propagation and an intrusion detection system checks the traffic between the network and the internet, the worm can easily infect the whole network because there is not a protection system that stops it. In this case, SANA'Õs artificial cells protect the nodes through distributed intrusion detection and disinfection of infected nodes. 

In the introduction, different criteria are introduced and SANA meets these:
\begin{itemize}
\item Due to the installation in all nodes SANA secures the whole network. All intrusions are not identified which is the ideal performance of a protection system. However, SANA identifies almost all intrusions. 
\item The system uses the resources efficient from all nodes. Thus, the required resources on a single node is reduced when SANA is used.
\item For the third criteria, the system should secure all nodes as well as it should not influence the normal production. These two criteria are met because SANA is installed on each node and uses only limited resources so that the normal production is not influenced. Furthermore due to its autonomous workflows SANA asks only rarely the user for critical security issues. 
\item The CNTS release regularly artificial cells, which check, repair, and update the different protection components for proper working - self-checking and -repairing. Furthermore, the installation is simplified because in each node must only the security environment installed and the protection components find a common infrastructure. Due to the enormous number of artificial cells is a breakdown of one cell not important for the overall performance.
\item SANA adapts to the current situation in the network. Therefore, a sophisticated information management is used and the protection components, i.e. artificial cells, identify abnormal behaviour, infected nodes, and adapt their behaviour to the current situation in the network.
\end{itemize}
After the practical implementation and analysis of the system, distributed protection systems are analysed more theoretically. Therefore, different attack scenarios are discussed where several attacks are always successful when current protection systems are used. Examples are that a user wants to attack a node, shuts it therefore down, and boots from an external storage device, e.g. Linux on an USB-stick. The protection components in this system are inactive and the user can install all intrusions and other protection components of current protection system does not identify it. SANA identifies infected nodes quickly, quarantines the node, and informs the administrator. Other attacks are that novel intrusions can infect a whole network because only centralised systems identify it. 

To sum up, distributed protection systems as SANA have lots of advantages compared to centralised collections of protection components, which are today widely used. However, there are lots of tasks to do in order to bring distributed protection systems into production and these are the challenges for the future in this research project. Two examples are implementation issues how the security environment should be installed in each node as well as the security problems so that an adversarial cannot use the distributed protection system in order to run attacks.

% =================================================================
\section{Conclusion}
\label{secConclusion}
% =================================================================
As it was described, the protection system SANA outperforms current protection systems. An intelligent administration and a distributed architecture with a standardised protection environment increases its performance. The distributed and dynamic framework makes it hard to attack and to break it completely down. Next challenges are to enlarge the library and to discuss the implementation of distributed protection system where the focus lays on the security problem that an adversarial can use the protection system for attacks. 

% =================================================================
\section*{Acknowledgments}
% =================================================================
SANA is currently implemented in the research project INTRA (= INternet TRAffic management and analysis) that takes place in the University of Luxembourg. We thank the Ministre Luxembourgeois de l'education et de la recherche for additional financial support. Furthermore, we thank Katja Luther, TU Berlin, for discussion and advise.

\bibliography{paper}

\begin{thebibliography}{10}

\bibitem{Aic04a}
U.~Aickelin, P.~Bentley, S.~Cayzer, J.~Kim, and JU. McLeod.
\newblock Danger theory: The link between ais and ids?
\newblock In {\em Proceedings of the Second International Conference on
  Artificial Immune Systems (ICARIS 2003). LNCS 2787, pp. 147-155}, Edinburgh,
  UK, 2003.

\bibitem{Aic04}
U.~Aickelin, J.~Greensmith, and J.~Twycross.
\newblock Immune system approaches to intrusion detection - a review.
\newblock {\em Proceedings of the 3rd International Conference on Artificial
  Immune Systems (ICARIS 2004)}, 2004.

\bibitem{Bel99}
F.~Bellifemine, A.~Poggi, and G.~Rimassa.
\newblock {JADE} - a {FIPA}-compliant agent framework.
\newblock In {\em Proceedings of the Practical Applications of Intelligent
  Agents}, 1999.

\bibitem{Deb99}
H.~Debar, M.~Dacier, and A.~Wespi.
\newblock Towards a taxonomy of intrusion-detection systems.
\newblock {\em Comput. Networks}, 31(9):805--822, 1999.

\bibitem{Dec02}
L.~N. DeCastro and J.~Timmis.
\newblock {\em Artificial Immune Systems: A New Computational Intelligence
  Approach}.
\newblock Springer, first edition, 2002.

\bibitem{Gre05}
J.~Greensmith, U.~Aickelin, and S.~Cayzer.
\newblock Introducing dendritic cells as a novel immune-inspired algorithm for
  anomaly detection.
\newblock {\em Proceedings of the 4th International Conference on Artificial
  Immune Systems (ICARIS 2005)}, 2005.

\bibitem{Hil06b}
M.~Hilker and C.~Schommer.
\newblock Agnosco - identification of infected nodes with artificial ant
  colonies.
\newblock {\em Proceedings of RASC}, 2006.

\bibitem{Hil06c}
M.~Hilker and C.~Schommer.
\newblock Sana - security analysis in internet traffic through artificial
  immune systems.
\newblock {\em Proceedings of the Trustworthy Software Workshop, Saarbruecken,
  Germany}, 2006.

\bibitem{Hof99}
S.~A. Hofmeyr and S.~Forrest.
\newblock Immunity by design: An artificial immune system.
\newblock {\em Proceedings of the Genetic and Evolutionary Computation
  Conference}, 2:1289--1296, 1999.

\bibitem{Hof00}
S.~A. Hofmeyr and S.~Forrest.
\newblock Architecture for an artificial immune system.
\newblock {\em Evolutionary Computation}, 8(4):443--473, 2000.

\bibitem{Jan04}
C.~A. Janeway, P.~Travers, M.~Walport, and M.~Shlomchik.
\newblock {\em Immunobiology: the Immune System in Health and Disease}.
\newblock Garland Publishing, sixth edition, 2004.

\bibitem{Mat02}
P.~Matzinger.
\newblock The danger model a renewed sense of self.
\newblock {\em Science}, 296(5566):301--305, 2002.

\bibitem{Roe99}
M.~Roesch.
\newblock Snort - lightweight intrusion detection for networks.
\newblock {\em LISA}, 13:229--238, 1999.

\bibitem{Var02}
P.~A. Vargas, L.~de~Castro, and F.~von Zuben.
\newblock Artificial immune systems as complex adaptive systems.
\newblock In {\em Proceedings of the 1st International Conference on Artificial
  Immune Systems (ICARIS)}, volume~1, pages 115--123, 2002.

\end{thebibliography}
\bibliographystyle{plain}
\end{document}